\documentclass[aps,prl,showpacs,notitlepage,superscriptaddress,floatfix,twocolumn]{revtex4}
\usepackage{bbm}
\usepackage{mathrsfs}
\usepackage{epsfig}
\usepackage{graphicx}
\usepackage{amsfonts}
\usepackage[figuresright]{rotating}
\usepackage{amssymb}
\usepackage{amsmath}
\usepackage{dcolumn}
\usepackage{bm}
\usepackage{braket}
\usepackage{comment}
\usepackage{mathtools}
\def\avg#1{\langle#1\rangle}

\def\be{\begin{equation}} \def\ee{\end{equation}}
\def\bea{\begin{eqnarray}} \def\eea{\end{eqnarray}}

\usepackage{color}

\usepackage[normalem]{ulem}

\begin{document}
\title{Interaction effects with varying $N$ in SU($N$) symmetric fermion lattice systems}
\author{Shenglong Xu}
\affiliation{Department of Physics, University of California,
San Diego, California 92093, USA}
\affiliation{
Condensed Matter Theory Center and Department of Physics, University of Maryland, College Park, Maryland 20742, USA}
\author{Julio T. Barreiro}
\affiliation{Department of Physics, University of California,
San Diego, California 92093, USA}
\author{Yu Wang}
\affiliation{
School of Physics and Technology, Wuhan University, Wuhan 430072, China}
\author{Congjun Wu}
\affiliation{Department of Physics, University of California,
San Diego, California 92093, USA}
\begin{abstract}
The interaction effects in ultracold Fermi gases with 
SU($N$) symmetry are studied non-perturbatively in half-filled one-dimensional lattices
by employing quantum Monte Carlo simulations.
We find that as $N$ increases, weak and strong interacting systems are driven
to a crossover region, but from opposite directions as a convergence
of itinerancy and Mottness.
In the weak interaction region, particles are nearly itinerant,
and inter-particle collisions are enhanced by $N$, resulting in the
amplification of interaction effects.
In contrast, in the strong coupling region, increasing $N$ softens the
Mott-insulating background through the enhanced virtual hopping processes.
The crossover region exhibits nearly $N$-independent physical quantities,
including the relative bandwidth, Fermi distribution, and the spin structure factor.
The difference between even-$N$ and odd-$N$ systems is most prominent
at small $N$'s with strong interactions, since the odd case allows
local real hopping with an energy scale much larger than the virtual one.
The above effects can be experimentally tested in ultracold atom experiments
with alkaline-earth (-like) fermions such as $^{87}$Sr ($^{173}$Yb).
\end{abstract}
\maketitle

High symmetry groups (e.g. SU($N$), Sp($N$), and SO($N$)) are typically
investigated in the context of high-energy physics.
They were introduced to condensed matter physics as a tool to apply the $1/N$-expansion to handle strong correlation effects in realistic SU(2)
electronic systems \cite{Affleck1988a,Arovas1988,Read1991,Sachdev1991}.
High symmetries enhance quantum spin fluctuations and suppress
antiferromagnetic N\'eel ordering \cite{Affleck1988a,Marston1989,
Arovas1988,Read1990}, which has stimulated intensive efforts to
study exotic quantum paramagnetic states, such as valence-bond
solid states and spin liquid states with high symmetries \cite{Paramekanti2007,Hermele2009,Honerkamp2004,
Harada2003,Corboz2011,Corboz2012,Lang2013a}.
However, high symmetries are rare in solids
in spite of many multi-component spin systems.
For example, in transition-metal-oxides, Hund's coupling aligns
electron spins forming large spin moments.
However, the symmetry remains SU(2), and quantum spin fluctuations are
suppressed by the large spin.

Ultracold atom systems open up new possibilities to study high symmetries
since many alkali and alkaline-earth fermions carry hyperfine spins larger
than $\frac{1}{2}$.
It was first pointed out that spin-$\frac{3}{2}$ fermionic systems exhibit
a generic Sp(4) symmetry without fine tuning \cite{Wu2003,WU2006,wu2005},
which is further enlarged to SU(4) for spin-independent interactions,
a feature of alkaline-earth fermions.
Magnetic and superfluid properties of large spin fermion systems,
many of which possess high symmetries, have been systematically studied
\cite{hoffman2017,capponi2016,he2015,guan2012,wangda2014,zhou2016a,ho2015,ho1999}.
The past few years have witnessed a significant experimental progress
along this direction.
High symmetries, such as SU(6) and SU(10), are realized with $^{173}$Yb \cite{Hara2011} and $^{87}$Sr atoms~\cite{Desalvo2010,pra-87-013611}, respectively.
The interplay between the nuclear-spin and electronic-orbital degrees
of freedom leads to complex physics \cite{Gorshkov2010,Hermele2009,Chen2016}.
Moreover, various SU($N$) symmetric quantum degenerate gases and Mott insulators in optical lattices have been realized~\cite{Gorshkov2010,Hara2011,
Taie2010,Scazza2014,Zhang2014a,Taie2012,Duarte2015,Desalvo2010,pra-87-013611,
Omran2015,Hofrichter2015,Pagano2014}.

A natural question on SU($N$) symmetric fermion systems is how fermion
correlations vary with $N$.
When the total fermion density is fixed, the Bethe-ansatz solution for
one-dimensional (1D) systems shows that the low energy properties
approach those of spinless bosons as $N\to \infty$ \cite{Yang2011b,guan2012}.
As $N$ becomes large, more and more fermions can
antisymmetrize their spin wavefunctions such that their orbital
wavefunctions are more symmetrized in mimicking bosons.
In a more interesting scenario, the density is fixed, but $N$ varies, as
realized in a recent experiment~\cite{Pagano2014}:
By using a subset of the spin-projection components of $^{173}$Yb atoms,
1D SU($N$) systems were constructed with up to $N=6$.
Increasing $N$ intensifies inter-particle collisions, and thus the Fermi
distribution is broadened as $N$ increases.
This experiment was performed in the metallic region where the
interaction effect is weak.

It would be interesting to explore further the consequences of a high symmetry
in the lattice, in particular, when the system is in Mott-insulating states,
for example, at half-filling.
In this letter, we systematically investigate the half-filled 1D SU($N$) lattice systems, i.e., $N/2$ particles per site on average,
by using quantum Monte-Carlo (QMC) simulations.
These systems are insulating in the ground states of both weak and strong
interaction regions, but the interaction effects scale differently
as $N$ increases when expressed in terms of the relative bandwidth
$W_R$ (the ratio of the kinetic energies in the interacting and free
systems), Fermi distribution, and spin structure factor.
In the weak interaction region, the interaction effects are strengthened
as $N$ increases.
In contrast, in the strong interaction region, increasing $N$ softens
the Mott-insulating background and weakens correlations.
There exists a crossover region characterized by nearly $N$-independent
relative bandwidths and spin structure factors.
To our knowledge, such a crossover phenomenon as the symmetry grows has not been investigated before.
Previous large-$N$ studies in the condensed matter literature typically focus on Heisenberg models by freezing the charge fluctuations, which are already in the
infinite-$U$ limit.
Our work is different from previous studies by directly working on
Hubbard models that include both charge and spin physics.
Due to the local charge fluctuations of odd-$N$ systems, they exhibit
an opposite $N$-dependence of the relative bandwidths and stronger
dimerization compared to the even-$N$ case at small values of $N$.
As $N$ increases, such difference is diminished by strong charge fluctuations.

We consider the 1D SU($N$) Hubbard model,
\bea
H=-t\sum\limits_{\langle ij \rangle,\alpha}c^{\dagger}_{i,\alpha}c_{j,\alpha}+
\frac{U}{2}\sum_i n_i(n_i-1) -\mu \sum\limits_{i}n_i,
\label{eq:hubbard}
\eea
where $\langle\rangle$ represents the nearest-neighboring bond;
the spin index $\alpha$ runs from 1
to $N$; $n_i=\sum\limits_{\alpha}c^{\dagger}_{i,\alpha}c^{}_{i,\alpha}$
is the total particle number at site $i$.
This model possesses a particle-hole symmetry at half-filling,
which fixes $\mu=\frac{N-1}{2} U$.

We will investigate quantitatively the correlation effects that
arise as $N$ varies from 2 to larger values,
by employing QMC, a method well-known to be free of the sign problem
in the path-integral framework in 1D at any filling.
The stochastic-series-expansion (SSE) QMC method will be applied
with the directed-loop algorithm \cite{Syljuasen2002}, which allows
us to perform large-scale simulations efficiently.
We will focus on insulating states at half-filling and
a system size set to $L=100$ for all simulations below.
The finite size effects were verified to be negligible
for all the quantities reported here.

\begin{figure}
\includegraphics[height=0.45\columnwidth, width=0.49\columnwidth]
{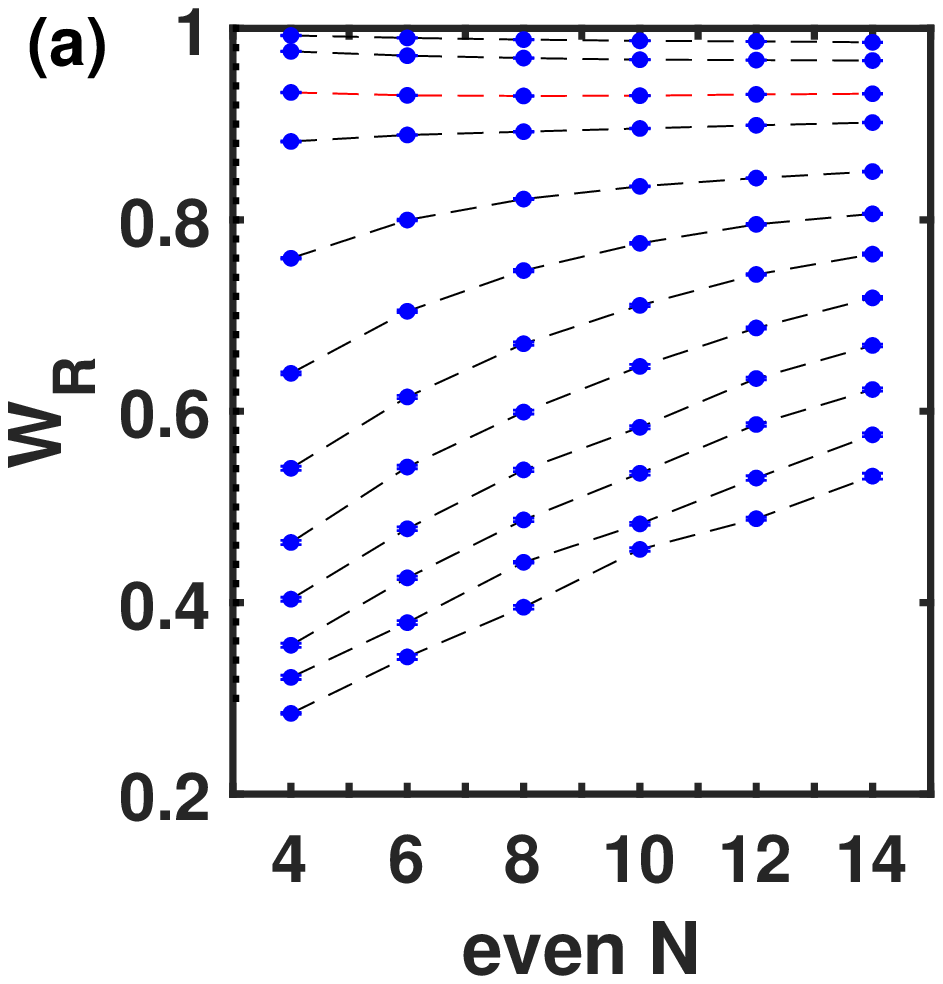}
\includegraphics[height=0.45\columnwidth, width=0.49\columnwidth]
{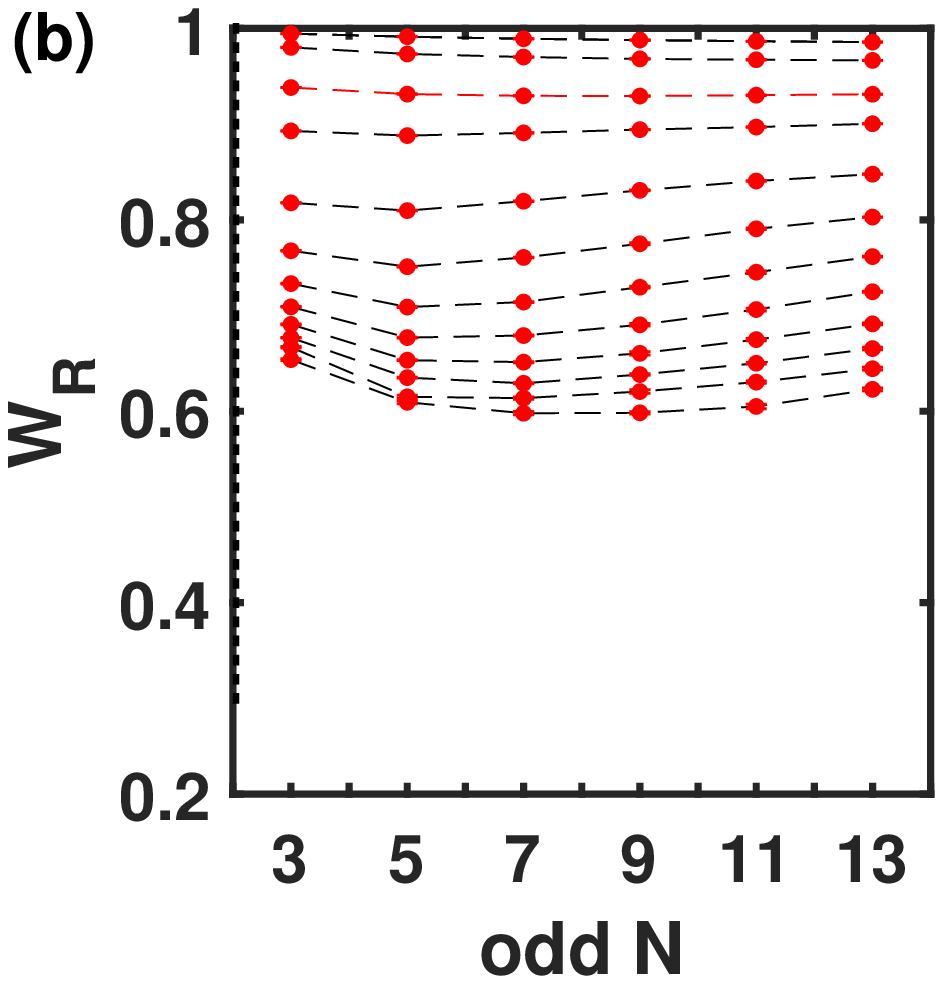}
\caption{The relative bandwidth $W_R$ for even
($a$) and odd ($b$) $N$ at $\beta=30$.
In both cases, the dashed lines shown as a guide from top to bottom correspond to
$U/t=0.5,1.0, 2.0, 3.0, 5.0, 7.0, 9.0, 11.0, 13.0, 15.0, 17.0, 19.0$, respectively.
The cross-over lines with $U/t\approx 2$ (marked red)
separating the weak
and strong interaction regions are nearly $N$-independent.
($a$) For even $N$, $W_R$ decreases with $N$ in the weak interaction region,
while it increases in the strong interaction region.
($b$) For odd $N$, the behavior for weak interactions is similar to ($a$). However, in the strong interaction regime, $W_R$ is non-monotonic,
first decreasing and then increasing with $N$.
}
\label{fig:kinetic}
\end{figure}

The ground states of the half-filled SU($N$) Hubbard chains described
by Eq. \ref{eq:hubbard} are insulating at $U>0$ with charge gaps
opening for all values of $N\ge 2$ \cite{assaraf2004,azaria2010,nonne2011,capponi2016}.
Except for the SU(2) case in which the spin sector remains gapless
exhibiting algebraic antiferromagnetic (AFM) ordering, spin gaps
open for $N\ge 3$ accompanied by the appearance of dimerization.
In the weak interaction region, there is no qualitative difference
between the even and odd $N$ cases.
Increasing $N$ enhances fermion collisions among different
components, which strengthens both charge and spin gaps.
However, in the strong interaction region $U\gg t$, the physics
is qualitatively different between even and odd $N$'s.
For illustration, consider a two-site problem filled with $N$ fermions,
which is discussed in detail in the Supplementary Material (SM.) I
\cite{supp}.
When $N$ is even, each site holds $\frac{N}{2}$ fermions on average.
Weak charge fluctuations arise from virtual hoppings,
generating the AFM super-exchange $J\approx 4t^2/U$.
In contrast, when $N$ is odd, the onsite charge fluctuations remain
significant even at the limit $U\to\infty$ due to the real hopping of one
particle in the background of $\frac{N-1}{2}$ particles on each site.
In both cases, the ground state is an SU($N$) singlet, and the first
excited states belong to the SU($N$) adjoint representation.
The spin gap for even values of $N$ is $\Delta_s\approx \frac{N}{2} J$,
while that for odd values of $N$ is $\Delta_s \approx t$.
The single-particle gap for adding one particle/hole is estimated as
$\Delta_{spg}\approx U$ when $N$ is even
and $\Delta_{spg}\approx \frac{N+1}{2} t$ when $N$ is odd.
Since dimerization develops in the 1D lattice for $N\ge 3$,
the picture based on two sites already captures the essential
physics in the thermodynamic limit.
As $N$ increases to $U/t$, the system crosses over into the weak interaction
region, and the distinction between even and odd $N$ cases smears.

\begin{figure}
\includegraphics[height=0.47\columnwidth, width=0.49\columnwidth]
{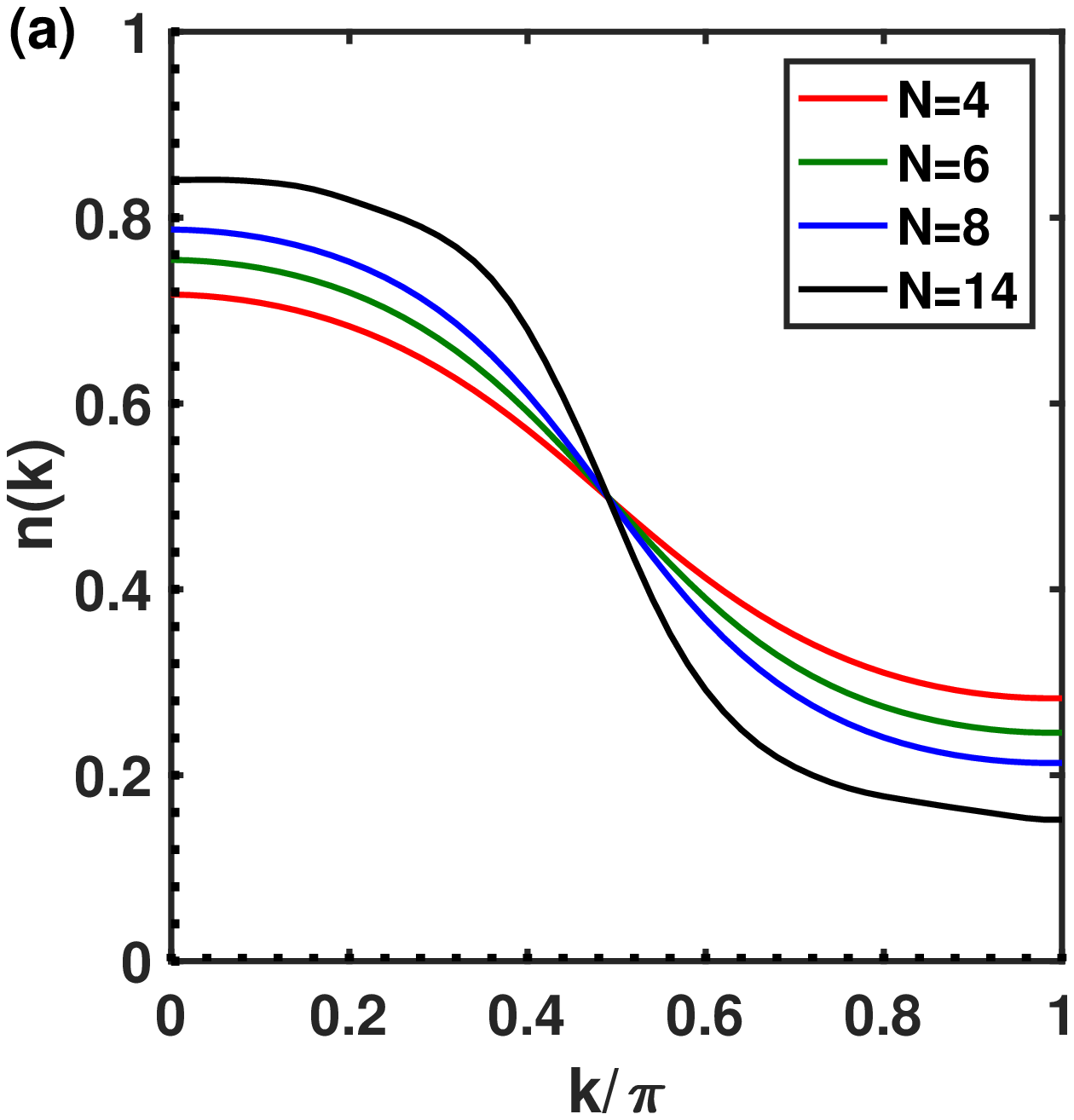}
\includegraphics[height=0.47\columnwidth, width=0.49\columnwidth]
{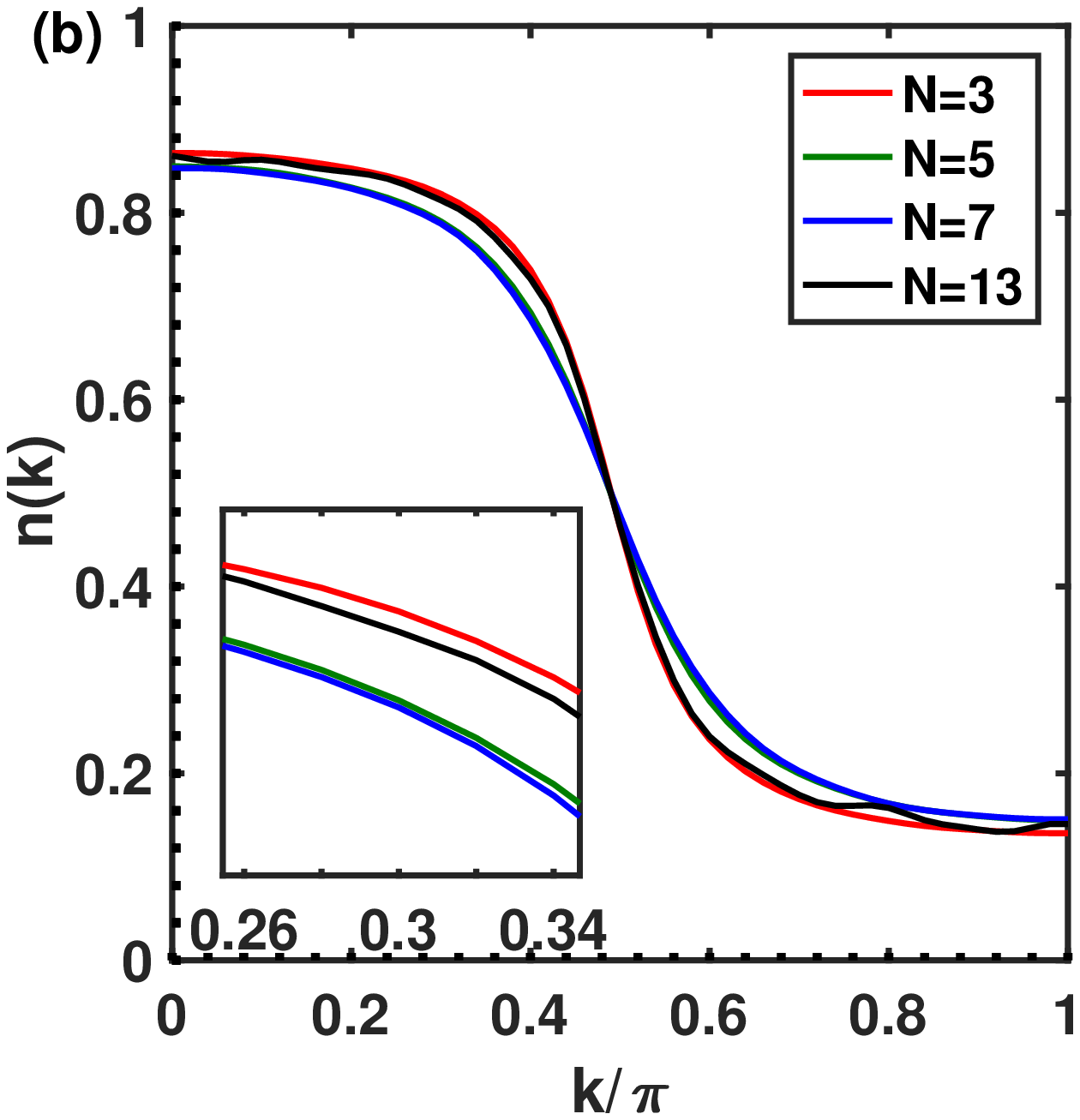}
\caption{Momentum distribution $n(k)$ in the strong
interaction region for even ($a$) and odd ($b$) values of $N$.
Parameter values are $U/t=15$ and $\beta=30$.
All curves cross at $n(k_f=\frac{\pi}{2})=\frac{1}{2}$.
($a$) For even $N$'s, $n(k)$ is driven towards the weak interaction
distribution limit as $N$ increases.
($b$) For odd $N$'s, $n(k)$ exhibits non-monotonic behavior as
$N$ increases, which is consistent with that of the $W_R$
shown in Fig.~\ref{fig:kinetic} ($b$).
}
\label{fig:nk}
\end{figure}

\begin{figure}
\includegraphics[height=0.47\columnwidth, width=0.49\columnwidth]
{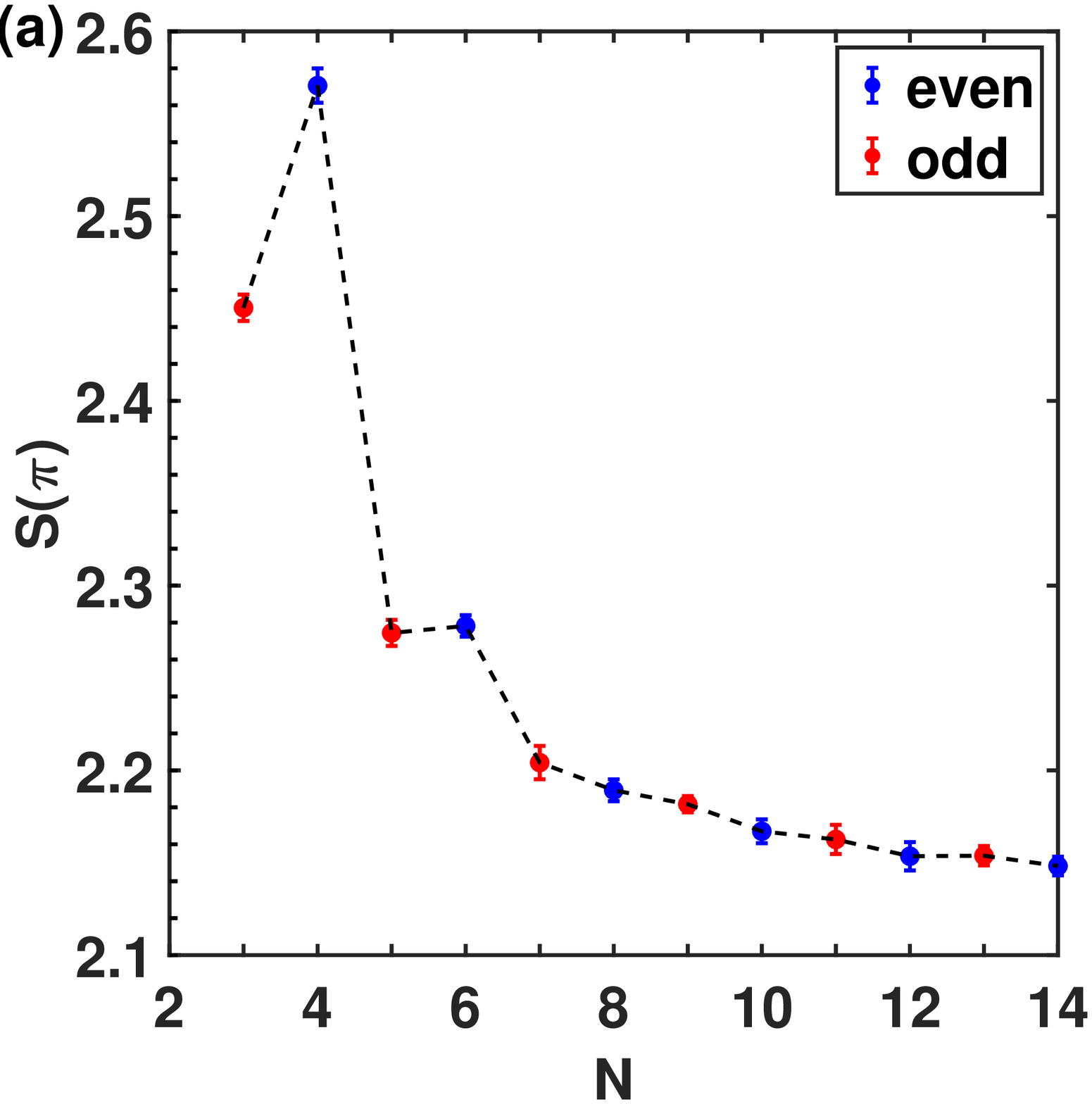}
\includegraphics[height=0.47\columnwidth, width=0.49\columnwidth]
{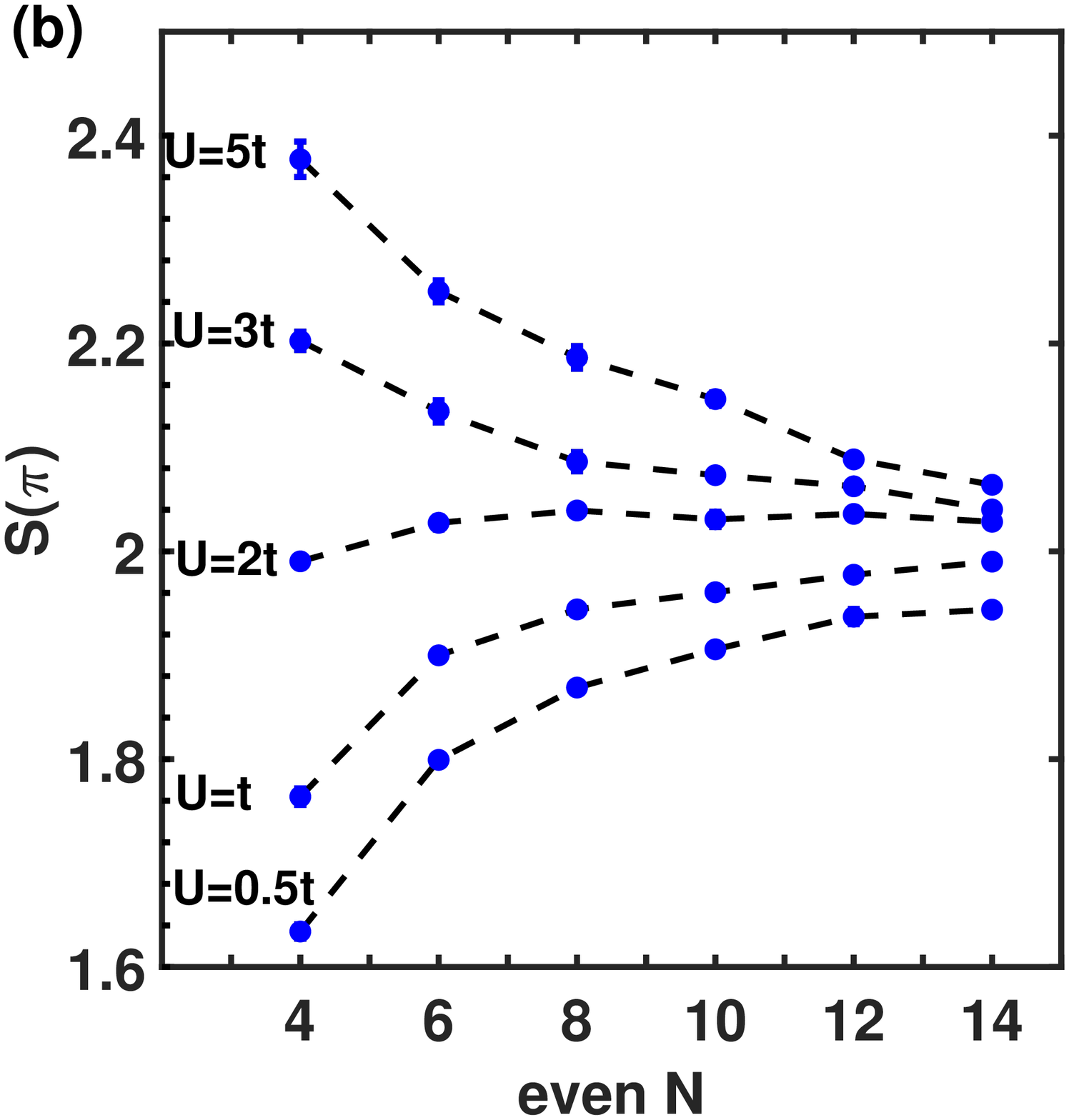}
\caption{ The $N$-dependence of $S(Q=\pi)$ at $\beta=30$.
  ($a$) At $U/t=15$ (strong interaction), $S(\pi)$ for $N=2m-1$ and $2m$
  are close, and the latter is slightly larger.
As $N$ increases, $S(\pi)$  drops rapidly showing the suppression of
AFM correlations.
($b$) Similar to $W_R$, $S(\pi)$ exhibits opposite $N$-dependence
in the weak and strong interaction regions.
The crossover occurs around $U/t\approx 2$,
consistent with $W_R$ shown in Fig.~\ref{fig:kinetic} ($a$).
}
\label{fig:spin}
\end{figure}

To support the above physical intuitions, we perform QMC simulations
at a very low temperature to approach the ground states.
($\beta=t/T=30$ is used for Figs. \ref{fig:kinetic}, \ref{fig:nk}, and \ref{fig:spin}.)
The bandwidth narrowing is a characteristic feature of Mott insulators,
which often shows in spectroscopy measurements in solids.
We define the relative bandwidth $W_R=E_K/E_K^0$ where $E_K$ is the kinetic
energy at the interaction $U$, and $E_K^0$ is the corresponding
non-interacting value.
$W_R=1$ in the non-interacting case, and it is completely suppressed
to zero at $U=\infty$.
At finite values of $U$, the $N$-dependence of $W_R$ from the weak to
strong interaction regions is plotted in Fig.~\ref{fig:kinetic} ($a$)
for even $N$ and ($b$) for odd $N$.
These curves do not cross since $W_R$ monotonically decreases as $U$ increases.
For small values of $U/t$, the single-particle gap $\Delta_{spg}$
is exponentially small, and thus fermions remain nearly itinerant over
a long correlation length $\xi \propto t/\Delta_{spg}$,
with the lattice constant set as 1.
Increasing $N$ strengthens the inter-particle collisions, and $W_R$
decreases monotonically
for both even and odd $N$.

Conversely, in the strong coupling region, a distinct even-odd effect
appears.
For $N$ even, $W_R$ is significantly below $1$ due to the suppression of
charge fluctuations.
The virtual hopping processes dominate, whose
number per bond scales as $(\frac{N}{2})^2$, thus $W_R$
increases roughly linearly with $N$.
For odd values of $N$, the overall scale of $W_R$ is larger than that
of even $N$'s, since both the real and virtual hopping processes
contribute to $W_R$.
When $N$ is small, the real hopping dominates, since its kinetic energy
scale is on the order of $t$, which is much larger than $J$ of the virtual
hopping.
Similar to the weak interaction case, $W_R$ goes down
initially as $N$ increases, which enhances inter-particle collisions.
As $N$ grows, the virtual hopping takes over,
since the number per bond scales as $(\frac{N-1}{2})^2$,
while that of the real hopping scales as  $\frac{N+1}{2}$.
Consequently, after passing a minima, $W_R$ increases with $N$.
The value of $N$ at the turning point can be determined by equaling
the energy scale of the real hopping $\frac{N+1}{2}t$ to
the virtual one $(\frac{N-1}{2})^2 J$ with $J=4t^2/U$,
yielding $N\approx U/2t$.
After that, the system crosses over from the strong to the intermediate
interaction region.
Then $W_R$ behaves similarly for both even and odd $N$.

Between the weak and strong interaction regions, there exist a crossover
area.
Say, along the line of $U/t\approx 2$, $W_R\approx 0.9$ is
nearly $N$-independent, as shown in both Fig.~\ref{fig:kinetic} ($a$) and ($b$).
For $N$ even, as $N$ increases, $W_R$ approaches the crossover
from opposite directions in the weak and strong interaction regions.
For $N$ odd, $W_R$ behaves similarly as $N$ becomes large.
This observation implies that the limits of $U\to 0$ and $N\to \infty$
are non-exchangeable.
For the non-interacting limit, $\lim_{N\to \infty} \lim_{U\to 0} (1- W_R)= 0$.
Moreover, we conjecture the existence of an interacting
large-$N$ limit
\bea
\lim_{U\to 0} \lim_{N\to \infty} (1-W_R) \approx 0.1.
\eea

The smearing of the Fermi distribution, which is defined as
$n(k)=\frac{1}{N}\sum_{\alpha} \avg{c^\dagger_{\alpha,k}c^{}_{\alpha,k}}$,
is an indication of correlation.
It can be measured via time-of-flight spectra in cold atom systems \cite{Pagano2014}.
Below we present the QMC results of the $N$-dependence of $n(k)$ in the strong
interaction region at a very low temperature for even and odd $N$'s
in Fig.~\ref{fig:nk} ($a$) and ($b$), respectively.
At $U=+\infty$, $n(k)$ is completely flattened.
Nevertheless, at finite values of $U$, short-range charge fluctuations
render $n(k_i)>n(k_f)>n(k_o)$,
where $k_i<k_f<k_o$ and the Fermi wavevector $k_f=\frac{\pi}{2}$.
Compared to the ideal Fermi distribution, $n(k)$ is significantly
smeared and becomes continuous at $k_f$.
Due to the particle-hole symmetry, $n(k)=1-n(\pi-k)$ holds for
all $N$'s at half-filling, thus all curves cross at $n(k_f)=\frac{1}{2}$.
When $N$ is even, $n(k)$ is less smeared as $N$ increases, which
enhances charge fluctuations on the Mott insulating background.
In contrast, the experiment done in 1D optical tubes observed
a broadening of $n(k)$, which is a feature of the weak-interaction metallic region \cite{Pagano2014}.
In our simulations, a similar behavior appears in the weak interaction
region where the fermion itinerancy remains significant.
When $N$ is odd, $n(k)$'s dependence on $N$ is much weaker.
It also exhibits non-monotonic behavior as $N$ increases, which is
consistent with $W_R$'s.

Quantum magnetic correlation is a fundamental property in the Mott insulating
states reflected by the spin structure factor $S(Q)$, which can be
measured through the noise correlations of the time-of-flight spectra \cite{Altman2004}.
$S(Q)$ is the Fourier component at momentum $Q$ of the two-point spin
correlation function $C_s(i,j)$ defined as
\be
C_s(i, j)=\frac{1}{2C(N)}\sum\limits_{\alpha\beta}\langle
S_{\alpha\beta}(r_i)S_{\beta\alpha}(r_j)\rangle,
\label{eq:spin_strfctr}
\ee
where $S_{\alpha\beta}(r_i)=c^\dagger_{i,\alpha}c_{i,\beta}-
\frac{n_i}{N}\delta_{\alpha\beta}$;
$C(N)$ is defined to normalize $C_s=1$ for $i=j$ in the limit of large $U$,
as shown in the SM. II \cite{supp}.
The antiferromagnetic correlation is reflected by $S(Q=\pi)$,
which is studied for varying $N$ and $U$ below.
In Fig.~\ref{fig:spin} ($a$), $S(Q)$ in the strong interaction
region ($U/t=15$) is presented.
Again consider a two-site problem in the large-$U$ limit for intuition:
The number of resonating spin configurations is $(2m)!/(m!)^2$
for both $N=2m-1$ and $2m$.
Consequently, $S(Q)$ for $N=2m-1$ and $2m$ are close to each other.
Nevertheless, the odd-$N$ case has a prominent dimerization tendency
facilitated by stronger charge fluctuations, thus $S(Q)$ at $N=2m$ is larger.
Overall, $S(Q)$ decreases rapidly with increasing $N$, which enhances
spin fluctuations.
In Fig.~\ref{fig:spin} ($b$), the $N$-dependence of $S(Q)$ is
presented for even $N$ by varying $U$.
Similar to $W_R$, there exists a crossover region, say, $S(Q)$ is
nearly independent on $N$ around $U/t\approx 2$, which is
approached in opposite directions from weak and strong interaction regions.
The $N$-dependence at odd $N$'s is similar
as presented in SM. II \cite{supp}.

\begin{figure}
\includegraphics[height=0.47\columnwidth, width=0.49\columnwidth]
{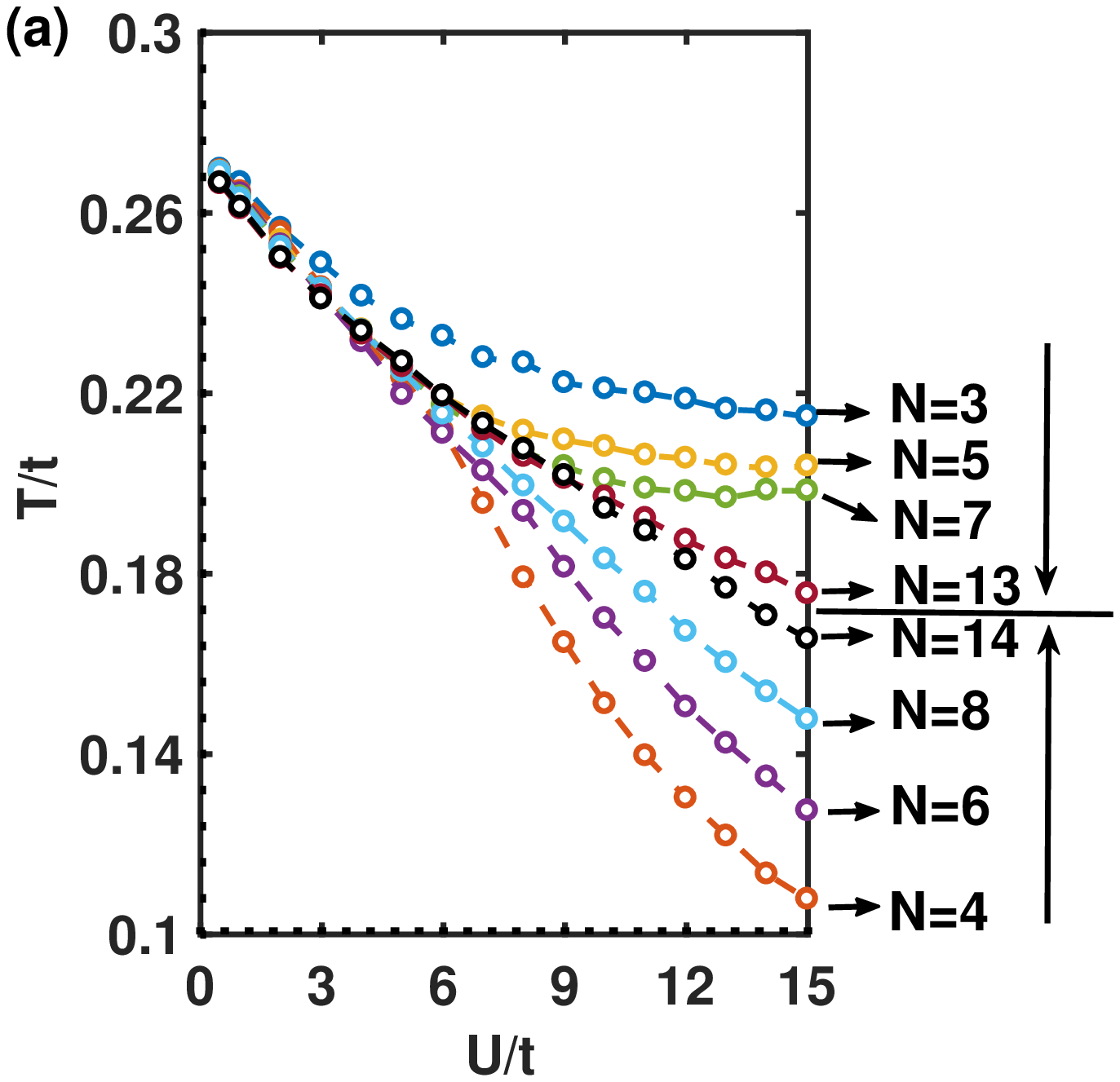}
\includegraphics[height=0.47\columnwidth, width=0.49\columnwidth]
{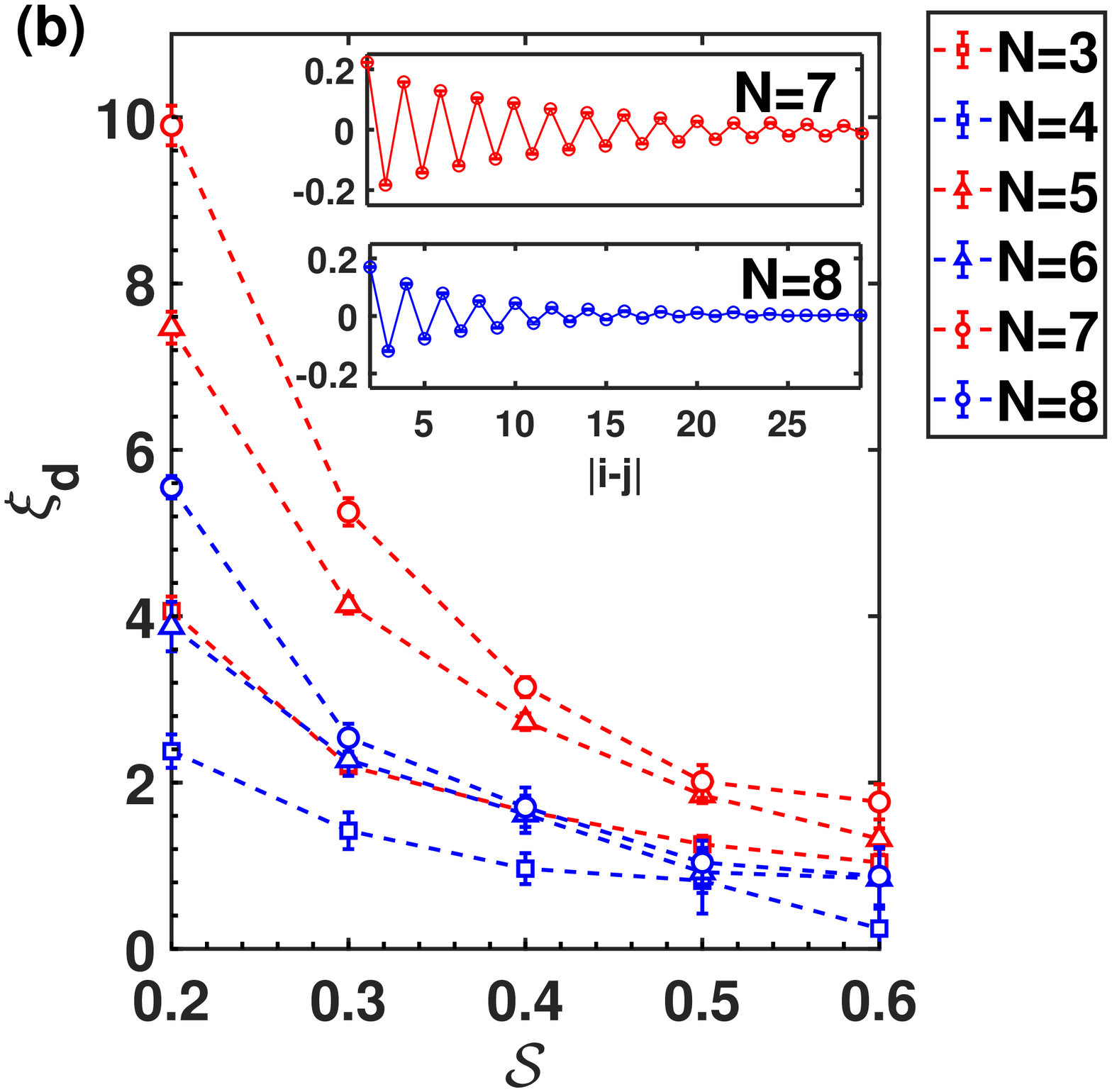}
\caption{($a$) The $T$-$U$ relations during the adiabatic process
with a fixed specific entropy $\mathcal{S}/k_B=0.3$.
As $U$ increases to the strong interaction region, the isoentropy
curves with even and odd $N$'s behave differently, and they
merge together from opposite directions in the large-$N$ limit.
($b$) The dimer correlation length $\xi_d$ v.s. the specific entropy $\mathcal{S}$.
The correlation length increases with $N$ in general.
The odd-$N$ systems (red) overall exhibit stronger dimerization instability
than the even-$N$ ones (blue).
In the inset, the real space correlation function $C_{d}(i,j)$
is shown for $N=7$ and $N=8$ for comparison at $\mathcal{S}/k_B=0.2$.
}
\label{fig:finite_entropy}
\end{figure}

\textit{Finite entropy-}
Now we consider the finite temperature (entropy) properties. In cold-atom lattice experiments, the entropy per particle $\mathcal{S}$, or specific entropy, is a more convenient parameter than temperature $T$.
The $\mathcal{S}(T)$ relation can be obtained via
\bea
\mathcal{S}(T)= \ln 4+\frac{E(T)}{T}-\int_T^\infty dT \frac{E(T)}{T^2},
\eea
where $\ln 4$ is the entropy at the infinite-$T$ limit and $E(T)$ is
the internal energy per particle.
We present the isoentropy curves in the $T$-$U$ plane as adiabatically
turning $U$ to the strong interaction region for different $N$'s
in Fig.~\ref{fig:finite_entropy} ($a$), considering $\mathcal{S}$ below, but
close to the experimental availability, say, $\mathcal{S}=0.3$~\cite{Omran2015}.
An overall trend is that $T$ decreases as $U$ increases for all $N$'s
as a reminiscence of the Pomeranchuk effect: Increasing $U$ drives the
system more local-moment like, and thus $T$ decreases to maintain $\mathcal{S}$
invariant \cite{Hazzard2012, Cai2013, Zhou2014,Zhou2016}.
The temperatures as $U$ becomes large for even $N$'s are lower than those
of odd $N$'s because fermions in the even $N$ case are more
local-moment-like and possess higher entropy capacity than the odd $N$ case.
When $N$ is even, as shown in previous studies \cite{Zhou2014,Cai2013},
increasing $N$ softens the Mott gap and drives the
system less local-moment-like, which reduces the entropy capacity,
and thus $T$ increases as $N$ increases.
In contrast, there are still significant local charge fluctuations
in the odd $N$ case even in the large $U$ region.
Increasing $N$ further enhances the collision among fermions,
and reduces the fermion itinerary, which increases the entropy
capacity and reduces the temperature.
This is in sharp contrast to the case of $1/N$-filling
investigated before, in which $T$ monotonically
decreases simply because of the $\ln N$ scaling of the specific
entropy \cite{Hazzard2012,Bonnes2012,prl-109-205306}.

The above even-odd effects can be observed experimentally by
  measuring the dimerization order.  When $N$ is odd, the
dimerization ordering is based on the combined effect of real and
virtual hoppings: The real hopping dominates if $N$ is small, and the
virtual one becomes important as $N$ goes large.  Hence, the
corresponding dimerization ordering is stronger than that of the even
$N$ case which is only based on the virtual hopping.  Therefore, we
propose that systems with odd $N$ are better candidates to observe
the dimerization order compared to those with even $N$.
In particular, systems with odd values of $N$, $3\le N \le 9$, are experimentally
accessible by using $^{173}$ Yb and $^{87}$Sr atoms
\cite{Hara2011,Desalvo2010,pra-87-013611}.
We define its two-bond correlation function as \bea
C_{d}(i-j) =\frac{1}{N} \left (\langle K_{i}K_{j}\rangle - \langle K_{i} \rangle \langle K_{j} \rangle \right ) ,
\eea
where
$K_i=\sum_\alpha c^\dagger_{\alpha, i} c_{\alpha, i+1} + h.c.$ is the
bonding strength between site $i$ and $i+1$.
 The dimer correlation
length $\xi_d$ is simulated by fitting $C_d(i-j)\propto e^{-|i-j|/\xi_d}$.
Its dependence on $\mathcal{S}$ and $N$ is plotted in
Fig.~\ref{fig:finite_entropy} ($b$) in the strong interaction
region.
As $\mathcal{S}$ decreases, $\xi_d$ grows much faster in the odd $N$ case.
For comparison,
$C_d(i-j)$'s are plotted in real space for $N=7$ and $8$ in the inset.

The lowest specific entropy $\mathcal{S}$ reachable in the optical lattice is
about at $\sim 0.6 k_B$, and in the center of the harmonic trap, it
can be lowered to $\sim 0.3 k_B$ \cite{Paiva2011, Omran2015}.
We expect that with further
improvements in cooling and spectroscopic
techniques~\cite{pra-86-023606,n-519-211,Taie2012,Zhang2014a}, the dimer ordering
could be observed in future SU($N$) cold atom experiments.
In particular, the following detection protocol would yield the two-bond correlation function above, but for alternating pairs of sites, and revealing its decay. After the realization of the 1D SU($N$) Hubbard model with $^{87}$Sr for any $N$ in a blue magic-wavelength lattice at 389.9~nm~\cite{pra-92-040501}, chosen for example and technical convenience, we follow the detection scheme in Refs.~\cite{s-340-1307,pra-92-040501}, but with a bichromatic lattice including one at 2$\times$389.9~nm. However, before band-mapping, when nearest-neighbor singlets-triplets lead to two-band occupation, we selectively~\cite{Zhang2014a} and sequentially transfer each band occupation and spin component to a state with a cycling fluorescing transition~\cite{prl-101-170504}, $^3$P$_2$, which is each then imaged with a quantum gas microscope~\cite{pra-91-063414,njp-18-23016}, ultimately revealing the dimer correlation length.

In conclusion, we have non-perturbatively studied one-dimensional SU($N$)
fermion lattice systems at half-filling.
In the strong interaction region, the odd-$N$ systems exhibit stronger charge fluctuations and dimerization than the even $N$.
As $N$ reaches the level of $U/t$, the virtual hopping processes
dominate in both even and odd $N$ systems, and the interaction effects
are weakened for increasing $N$.
Whereas from the weak interaction limit, increasing $N$ enhances particle
collisions and strengthens the interaction effect.
These two distinct behaviors approach a crossover region around $U\sim 2t$
from opposite directions, as demonstrated in experimentally measurable
quantities including the kinetic energy scale, the momentum distribution
functions, and spin structure factors.
The above pictures of convergence of physics of itineracy and Mottness
are not limited to one dimension.
It applies to Mott states in two and three
dimensions as well.
In previous simulations of 2D SU($2N$)Hubbard models \cite{Zhou2016,Cai2013a}
the softening of the single particle gap for increasing $N$ has been
found in at relatively large values of $U$.
A detailed study will be deferred to a future letter.

{\it Acknowledgments}
S. X. and C. W. are supported by AFOSR FA9550-14-1-0168.
Y.W. gratefully acknowledges financial support from the National Natural Science
Foundation of China under Grants No. 11729402, No. 11574238, and No. 11328403. Y.W. is also grateful for the award of a scholarship funded by the China Scholarship Council (File No. 201706275082).
C. W. and J. B. acknowledge support from the President's Research
Catalyst Awards of the University of California.


\end{document}


\title{Supplemental Material: Interaction effects from the parity of $N$ in SU($N$) symmetric fermion lattice systems}
\author{Shenglong Xu}
\affiliation{Department of Physics, University of California,
San Diego, California 92093, USA}
\author{Julio Barreiro}
\affiliation{Department of Physics, University of California,
San Diego, California 92093, USA}
\author{Yu Wang}
\affiliation{
School of Physics and Technology, Wuhan University, Wuhan 430072, China}
\author{Congjun Wu}
\affiliation{Department of Physics, University of California,
San Diego, California 92093, USA}
\begin{abstract}
We present a few details supporting the conclusions in the main text, including estimating the single particle gap $\Delta_{spg}$ and the spin gap $\Delta_{s}$ based on the two-site picture for both even and odd $N$ cases, the
density-matrix-renormalization-group (DMRG) calculation of the gaps
for the SU(3) case, and the discussion on the SU($N$) spin correlations.
\end{abstract}
\maketitle

\section{The two-site problem}
In this section, we solve the two-site problem in the large-$U$ limit
for both even and odd values of $N$ for intuitions.
The corresponding single-particle gaps and the spin gaps are calculated.

\begin{figure}
\includegraphics[height=0.6\columnwidth, width=0.6\columnwidth]
{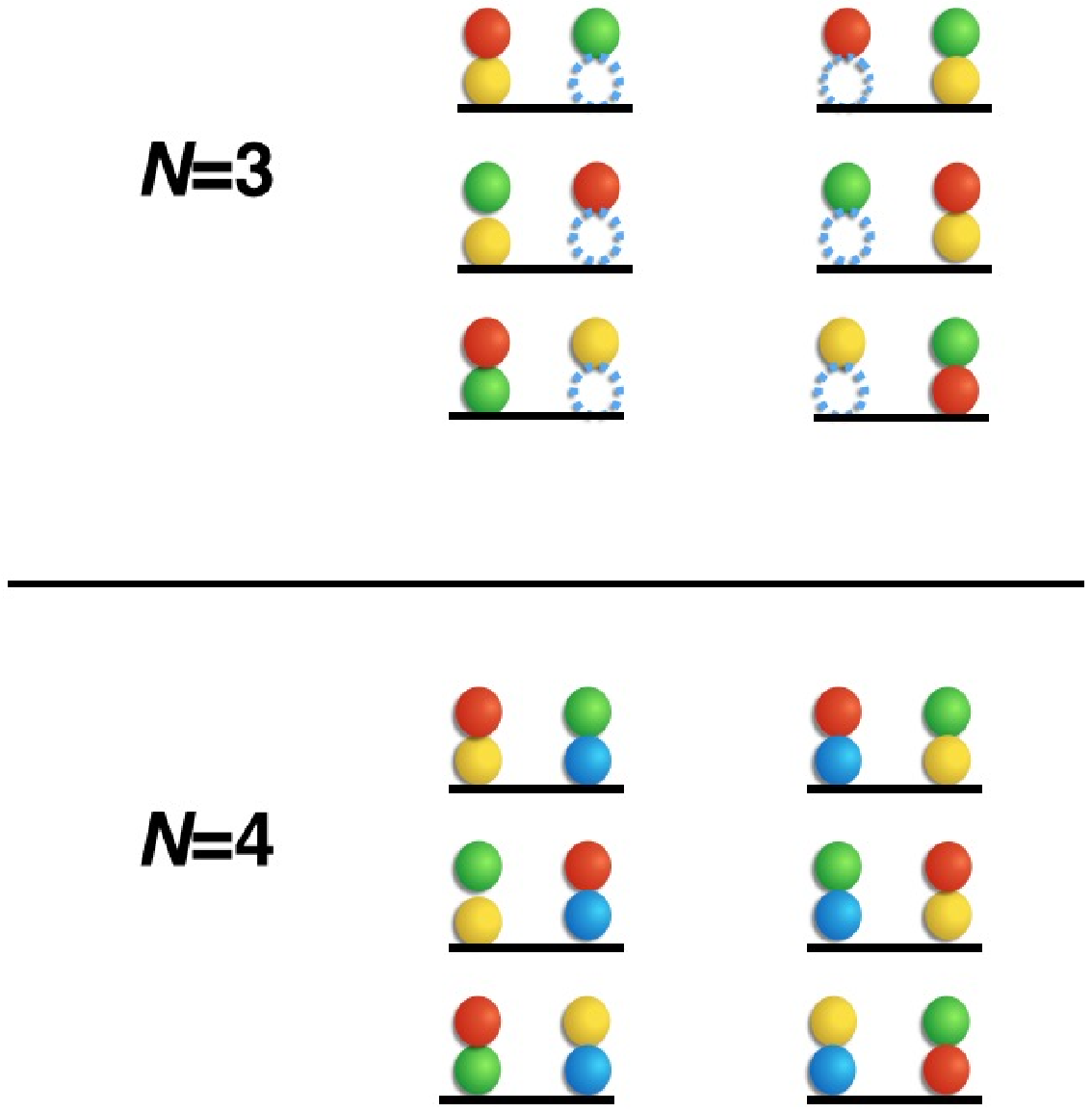}
\caption{The resonating spin configurations for a two-site SU($N$)
singlet in the large-$U$ limit.
The numbers of spin configurations are the same for $N=2m$ and
$N=2m-1$.
Examples of $N=3$ and $N=4$ are shown.
}
\label{fig:dimer_sketch}
\end{figure}

In the case of even $N=2m$, each site is filled with $m$ fermions in the
large-$U$ limit.
The single particle gap is simply $\Delta_{spg}=\frac{U}{2}$.
The dimensions of the truncated Hilbert space of this two-site problem
is $\left ( \frac{(2m)!}{m!m!} \right )^2$.
The degeneracy of these states is lifted by considering virtual hopping
to high energy states.
By performing the 2nd order perturbation theory, the effective
Hamiltonian within the truncated Hilbert space is
\bea
H_{eff}^e= \frac{J}{2} \left (\sum\limits_{\alpha,\beta} S_{\alpha\beta}(1)S_{\beta\alpha}(2) -\frac{N}{4} \right).
\label{eq:Heff_even}
\eea
with $J=\frac{4t^2}{U}$ is the super-exchange energy scale.
Its ground state is an SU($N$) singlet, denoted as $\ket{G}$, satisfying,
\bea
\Big( S_{\alpha\beta}(1)+S_{\alpha\beta}(2)\Big)\ket{G}=0.
\label{eq:singlet}
\eea
Therefore the ground state energy of Eq. (\ref{eq:Heff_even}) is,
\bea
E_{g}^e(N)&=& -\frac{J}{2} \left (\sum\limits_{\alpha,\beta}\bra{G} S_{\alpha\beta}(1)S_{\beta\alpha}(1) \ket{G}+\frac{N}{4} \right )
\\ \nn  &=&-\frac{J}{8}N(N+2).
\eea
The ground state contains all $N$ fermion components with $\frac{(2m)!}{(m!)^2}$
resonating spin configurations, as shown in Fig.~\ref{fig:dimer_sketch}.
The first excited state belongs to the SU($N$) adjoint representation.
To obtain the spin gap, one can change one fermion component and calculate
the energy difference.
Then there are two fermions with the same favor on different sites,
which do not contribute to the virtual hopping process.
Hence, effectively, the fermion favor numbers are reduced to $N-2$.
Therefore, the lowest energy in this sector is the same as the ground state
energy $E_g$ for the case of $N'=N-2$.
In consequence, the spin gap $\Delta_s=E_{g}^e(N-2)-E_{g}^e(N)=\frac{J}{2}N$.

In the case of odd $N=2m-1$, each site contains either $m$ or $m-1$ fermions
in the large-$U$ limit at half-filling.
The dimensions of the Hilbert space are still
$\left(\frac{(2m)!}{m! m!} \right )^2$.
Unlike the even-$N$ case, these states are connected through the real hopping.
The effective Hamiltonian is
\be
H_{eff, o}=-tP\left ( \sum_{\alpha}c^\dagger_{\alpha, 1} c_{\alpha, 2} \right )
P+h.c. ,
\ee
where $P$ is the projection operator into the above physical Hilbert space.
The ground state is also an SU($N$) singlet.
With the filling constraint, one can construct two singlet states $\ket{G}_a$
and $\ket{G}_b$.
The former has $m-1$ and $m$ fermions on site 1 and 2, respectively,
and the latter switches the occupation numbers.
The ground state is the superposition of these two as
\bea
\ket{G}=\frac{\sqrt{2}}{2}(\ket{G}_a+\ket{G}_b).
\eea
The number of the resonating spin configurations is still in total $\frac{(2m)!}{(m!)^2}$ as the same as the case of $N=2m$ as shown in Fig.~\ref{fig:dimer_sketch}.
To calculate the ground state energy, actually it is easier to compute
the expectation value of $H^2$ with respect to $\avg{H^2}$, and then
take the square root.
The expression of $(H_{eff}^0)^2$ is
\be
(H_{eff}^0)^2= -t^2 \left ( \sum\limits_{\alpha,\beta}S_{\alpha\beta}(1)S_{\beta\alpha}(2)-\frac{(N+1)^2}{4N} \right).
\ee
Therefore,
\bea
E_{g}^o &=&-\sqrt{\bra{G}(H_{eff}^o)^2\ket{G}}\nn \\
&=&-t \left ( \sum\limits_{\alpha,\beta} -\bra{G}S_{\alpha\beta}(1)S_{\beta\alpha}(1)\ket{G}+\frac{(N+1)^2}{4N} \right)^{1/2}\nn \\
&=&-\frac{N+1}{2}t.
\eea
For the second line, we have used Eq. (\ref{eq:singlet}) again.

We now discuss the single particle gap and the spin gap of the odd-$N$ systems.
After adding an extra fermion to the system, the real hopping processes are completely
forbidden by the filling constraint imposed by the strong Hubbard $U$, and
the energy is zero. Therefore, the single particle gap $\Delta_{spg}=\frac{t}{2}(N+1)$.
Similar to the even-$N$ case, the spin gap can be obtained by changing one fermion's flavor and calculate the energy difference, then the two fermions
with the same flavor occupying different sites do not participate in
hopping processes.
Hence, the lowest energy in this sector is the ground state energy for $N'=N-2$,
Consequently, the spin gap can be obtained as $\Delta_{s}=E_{g}^o(N-2)-E_{g}^o(N)=t$.

The single-particle and the spin gaps of the two-site problem are
summarized as below,
\bea
\label{eq:gap}
\begin{tabular}{| c | c | c| }
    \hline
    $N$ & even & odd \\ \hline
    $\Delta_{spg}$ & $\frac{U}{2}$ & $\frac{t}{2}(N+1)$ \\[5pt] \hline
    $\Delta_{s}$ & $\frac{J}{2} N$ & $t$ \\[5pt]
    \hline
  \end{tabular}
\eea

\begin{figure}
\includegraphics[height=0.45\columnwidth, width=0.45\columnwidth]
{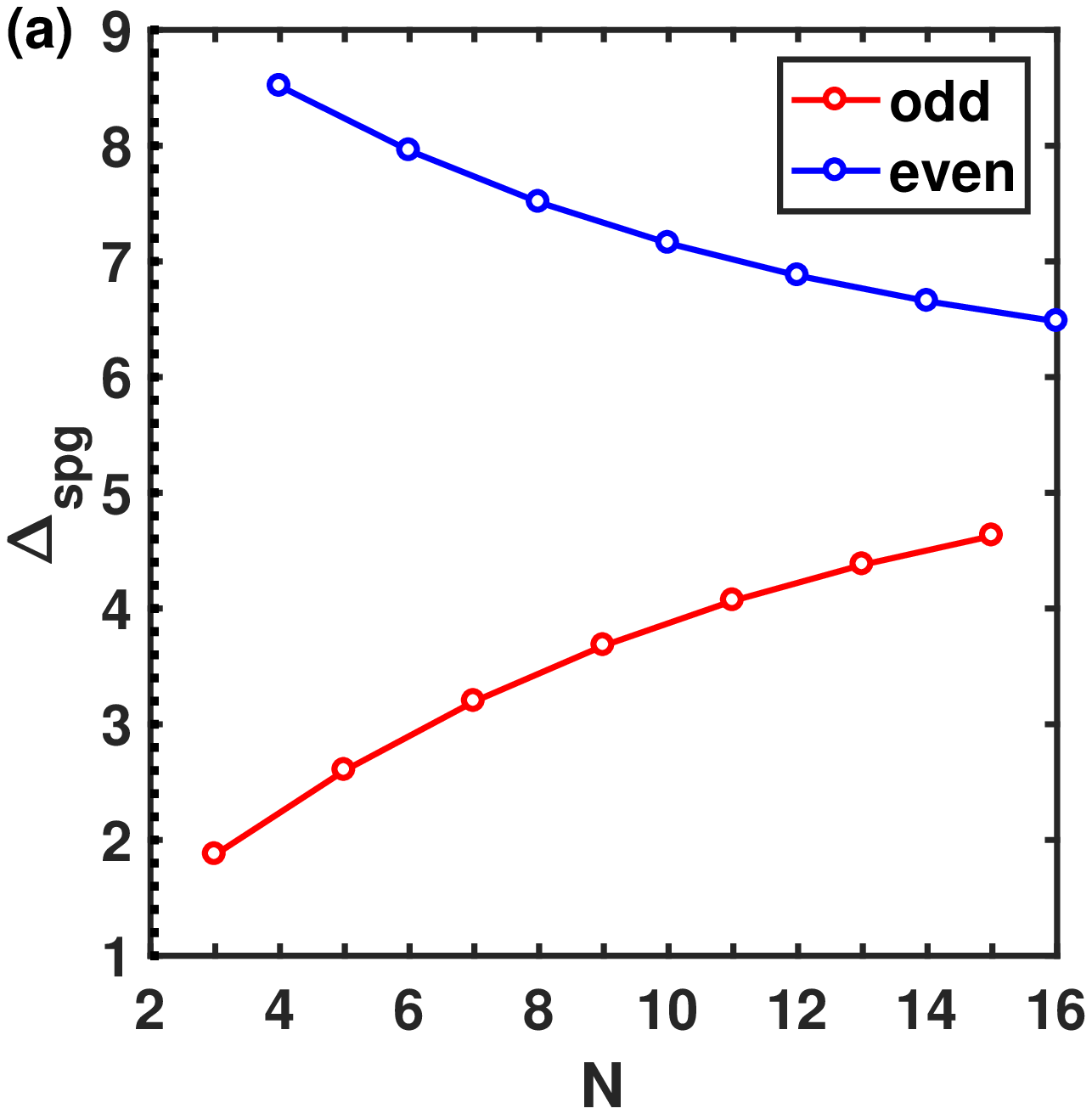}
\includegraphics[height=0.45\columnwidth, width=0.45\columnwidth]
{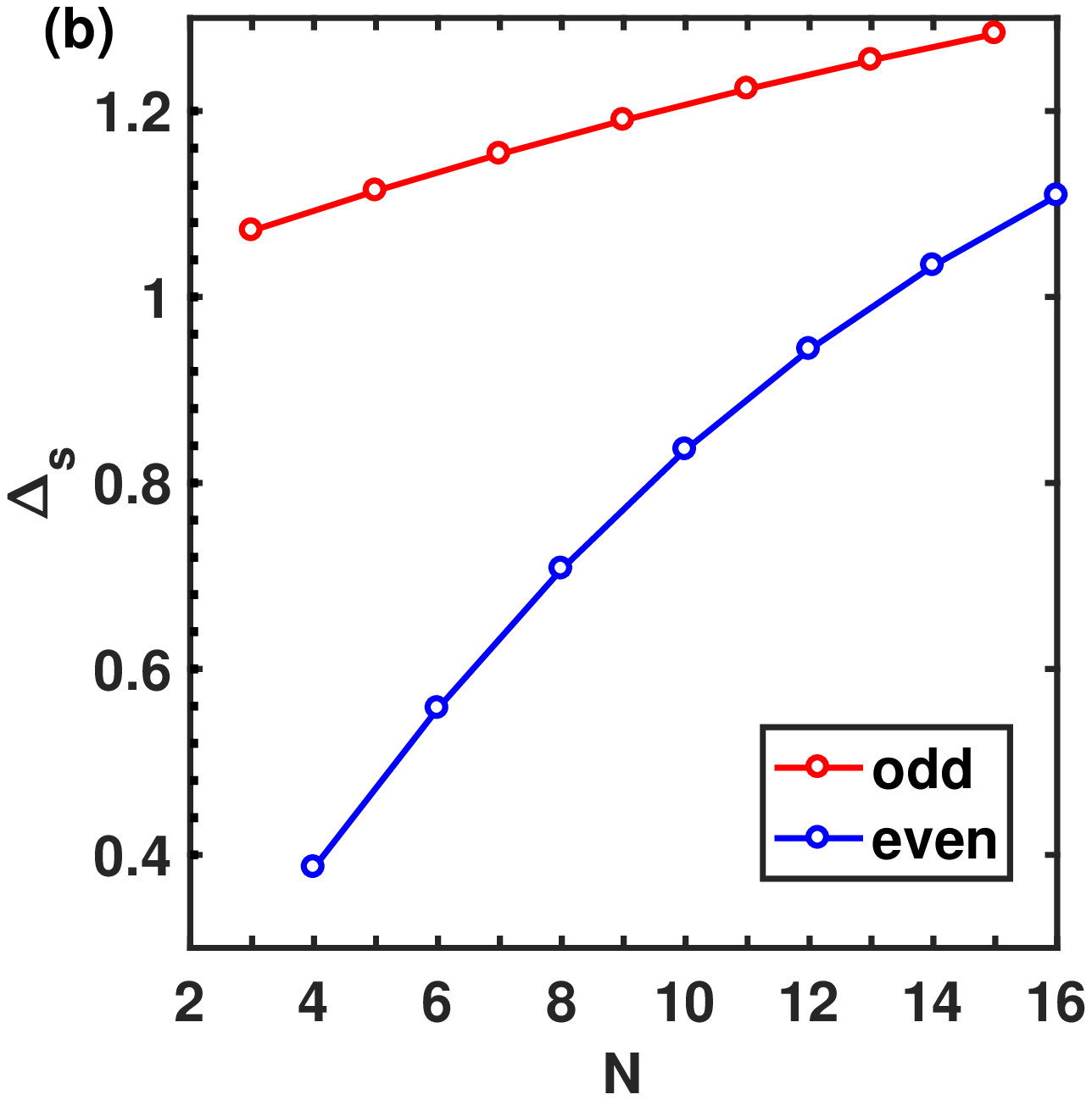}
\caption{The $N$-dependence of the single particle gap $\Delta_{spg}$
($a$) and the spin gap $\Delta_s$ ($b$) of the two-site problem.
The parameter value is $U/t=20$.
As $N$ increases, the different energy scales of $\Delta_{spg}$ and
$\Delta_s$ between even and odd-$N$ cases gradually merge together.  }
\label{fig:gap}
\end{figure}

The above analysis is valid for the strong interaction region $U/t \gg N$,
where the even-$N$ systems and odd-$N$ systems exhibit distinct
energy scales.
We also consider the 2-site problem with a fixed large value of $U$, say, at $U/t=10$ by using exact diagonalization.
As increasing $N$, the differences between even and odd values of $N$
diminish.
The two kinds of systems approach to the same limit, as shown in Fig.~\ref{fig:gap}, where the $N$-dependences of both
$\Delta_{spg}$ and $\Delta_{s}$ are presented.

\section{The DMRG calculations for the SU(3) chain}

\begin{figure}
\includegraphics[height=0.6\columnwidth, width=0.6\columnwidth]
{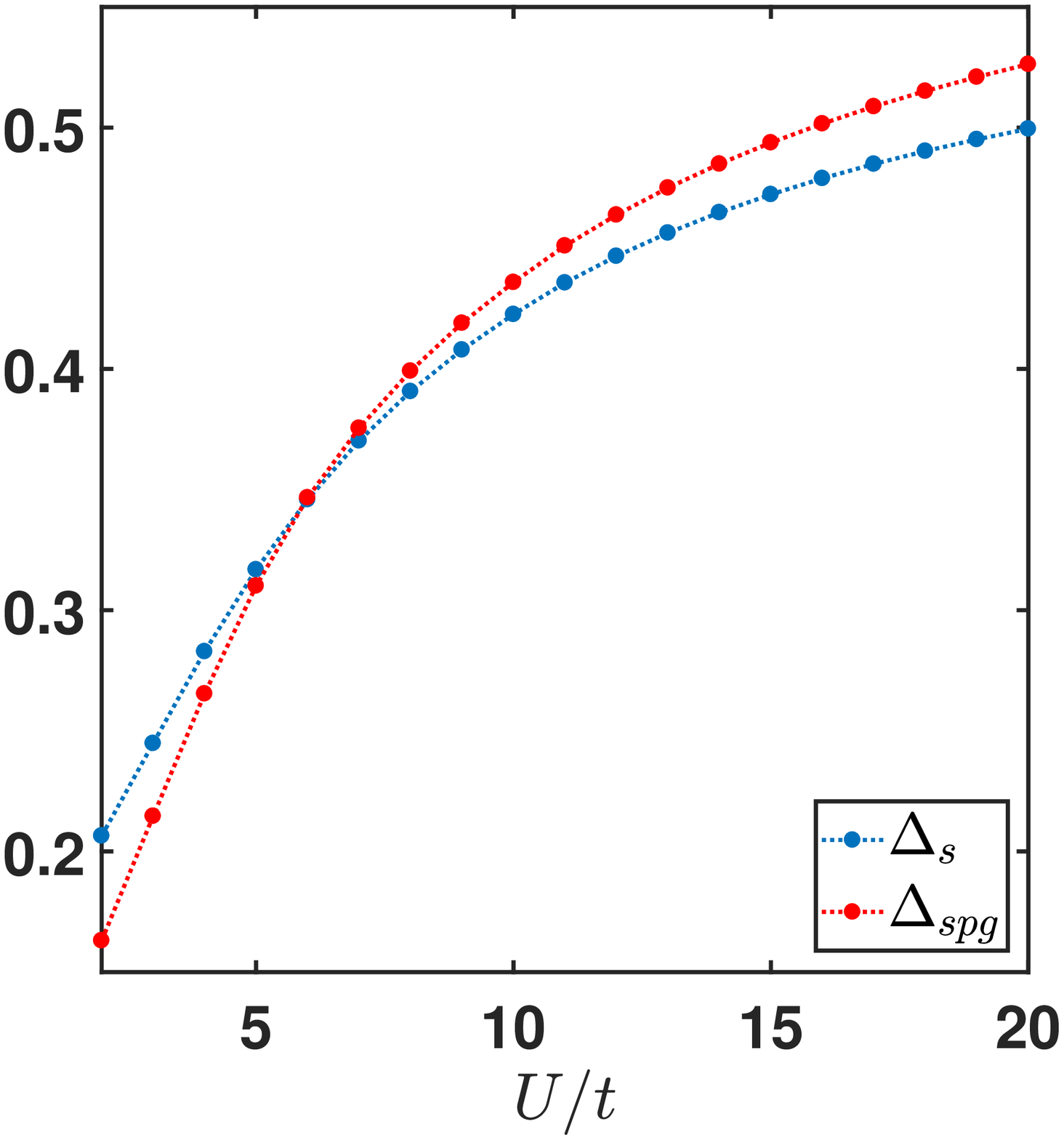}
\caption{The $U$-dependence of the single particle gap $\Delta_{spg}$
and the spin gap $\Delta_s$ for $N=3$ based on the DMRG calculations
for a lattice with 100 sites and the bond dimension 128.
In sharp contrast to the even-$N$ cases, both $\Delta_{spg}$ and
$\Delta_s$ approach finite values at the order of $\sim 0.5t$
as $U$ goes large, consistent with the two-site picture.}
\label{fig:su3}
\end{figure}

To justify the above argument, we calculate both $\Delta_{spg}$ and
$\Delta_{s}$ for a 100-site SU(3) Hubbard chain at half-filling, using
the DMRG method with bond-dimension 128, and the results are shown
in Fig. \ref{fig:su3}.
In the large $U$ limit, both the single-particle gap and the spin gap
approach $\sim 0.5 t$, drastically different from the even-$N$ case where $\Delta_{spg}\sim U/2$ and $\Delta_{s}\sim JN/2$.
Compared with two-site picture, the spin gap is reduced to half by
the many-body quantum fluctuation.

\section{The SU($N$) spin correlation function }
\label{spin_strfctr}
In this section, we provide some details on the spin correlation function
defined in the main text,
\bea
C_s(i, j)=\frac{1}{2C(N)}\sum\limits_{\alpha\beta}\langle
S_{\alpha\beta}(r_i)S_{\beta\alpha}(r_j)\rangle,
\label{eq:spin_strfctr}
\eea
where $S_{\alpha\beta}(r_i)=c^\dagger_{i,\alpha}c_{i,\beta}-
\frac{n_i}{N}\delta_{\alpha\beta}$. The normalization factor $C(N)$ is used such that $C_s(i,i)\to 1$
as $U\to\infty$.
It is straightforward to show that,
\be
C_s(i, i)=\frac{1}{2C(N)}\frac{N+1}{N}(N-n_i)n_i,
\ee
where $n_i$ is the on-site particle number operator.
In the case of even $N$, $n_i=N/2$, whereas in the case of odd $N$,
$n_i=(N\pm 1)/2$.
Requiring that $C_s(i,i)=1$, we have
\be
C(N) = \left\{ \begin{array}{ll}
\frac{N(N+1)}{8} & \mbox{if $N$ is even};\\
\frac{(N+1)(N^2-1)}{8N}& \mbox{if $N$ is odd}.\end{array} \right.
\ee

In the framework of the SSE Monte Carlo, measuring the off-diagonal
operators involves breaking world-lines and tracking relative
positions of the worm-head and tail, and thus is quite complicated.
Instead, the diagonal spin correlation function is more convenient to
measure\cite{Manmana2011}:
\bea
\braket{C_{s}^{diag}(i,j)}=\frac{1}{N}\sum\limits_{\alpha}\braket{n_i^\alpha n_j^\alpha}-\frac{1}{N(N-1)}\sum\limits_{\alpha\neq\beta}
\braket{n_i^\alpha n_j^\beta}, \nn
\label{eq:spin_corr_diag}
\eea
where $\braket{...}$ represents the thermal average.
In fact, one can show that the SU($N$) symmetry implies that
\be
\braket{C_s(i,j)}=\frac{N^2-1}{2C(N)}\braket{C_{s}^{diag}(i,j)}.
\nn \\
\ee
Therefore, the two kinds of correlation functions are equivalent
up to an overall constant.

\begin{figure}
\includegraphics[height=0.5\columnwidth, width=0.5\columnwidth]
{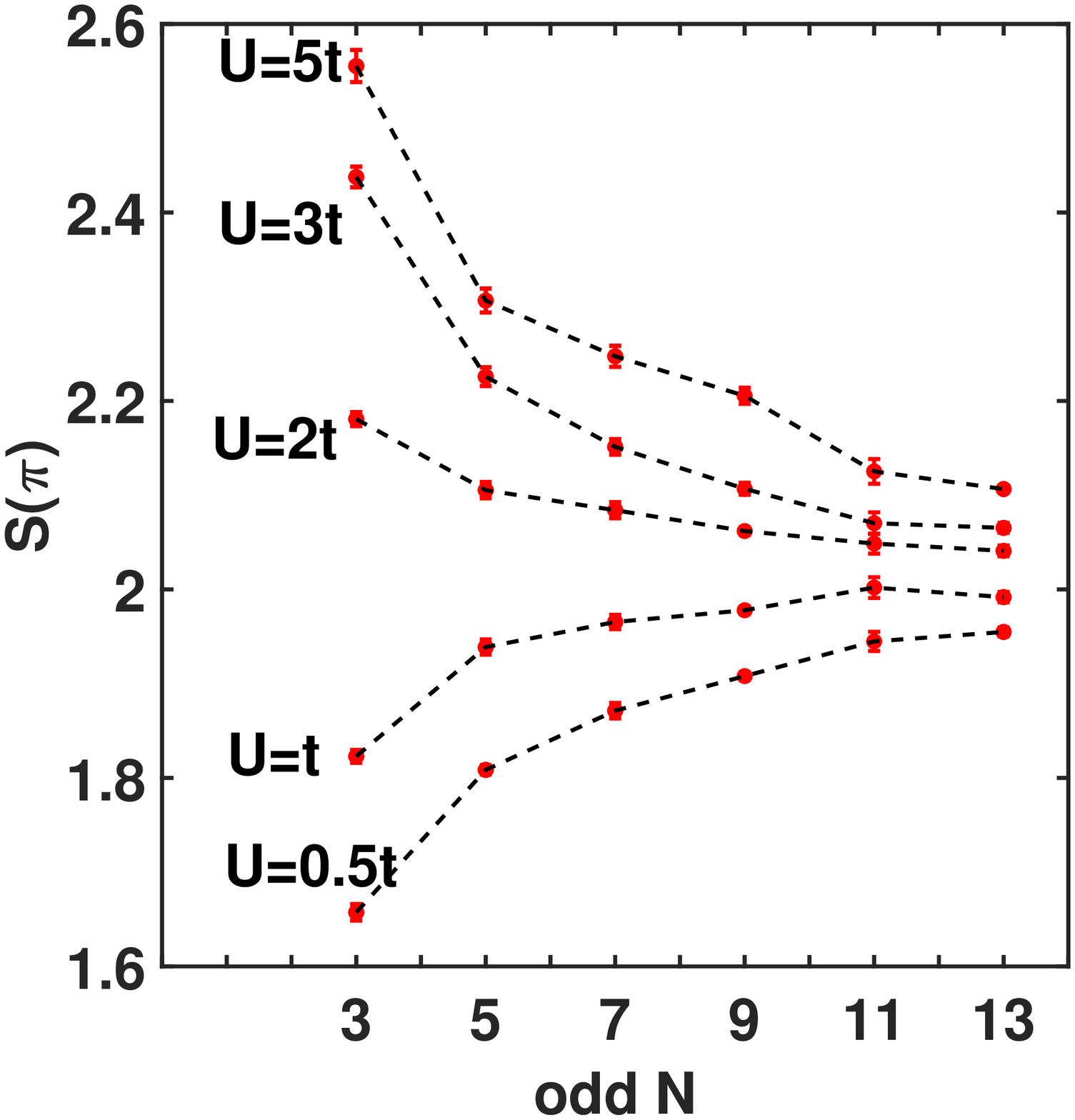}
\caption{The $N$-dependence of the spin structure factor $S(Q=\pi)$ for odd $N$. It is similar to the case of even $N$ presented in the main text. }
\label{fig:spin_odd}
\end{figure}

In Fig 3. (b) of the main text, we have presented the spin structure factor
for the case of even $N$, which shows a crossover behavior between the weak
interaction region and the strong interaction region.
Here, we present the spin structure factors of odd-$N$ systems
in Fig.~\ref{fig:spin_odd}, which show qualitatively the same
behavior as the even-$N$ systems.

